\documentclass[]{interact}

\input epsf
\usepackage{graphicx}
\usepackage{xcolor}
\usepackage{bm}
\usepackage{graphicx,float, pifont}
\usepackage{subfigure}

\def\R{\mathbb R}    

\def\pmb#1{\setbox0=\hbox{#1}
\kern-.025em\copy0\kern-\wd0 \kern-.05em\copy0\kern-\wd0
\kern-.025em\raise.0433em\box0}
\def\bg {\color{black}}
\def\bla {\big\langle}
\def\bra {\big\rangle}

\newcommand{\beq}{\begin{equation}}
\newcommand{\eeq}{\end{equation}}
\newcommand{\ba}{\begin{eqnarray}}
\newcommand{\ea}{\end{eqnarray}}

\input epsf

\usepackage[english]{babel}

\usepackage{epstopdf}
\usepackage[caption=false]{subfig}

\usepackage[numbers,sort&compress]{natbib}
\bibpunct[,]{[}{]}{,}{n}{,}{,}
\makeatletter
\def\NAT@def@citea{\def\@citea{\NAT@separator}}
\makeatother

\theoremstyle{plain}
\newtheorem{theorem}{Theorem}[section]
\newtheorem{lemma}[theorem]{Lemma}

\theoremstyle{definition}

\theoremstyle{remark}
\newtheorem{remark}{Remark}

\begin{document}

\articletype{ARTICLE}

\title{Magneto-electric coupling and magnetism in photonic crystals: A high-order homogenization approach in the constitutive equations.}

\author{
\name{B. Gralak\textsuperscript{a}, Y. Liu\textsuperscript{b} and S. Guenneau\textsuperscript{c}\thanks{CONTACT S. Guenneau. Email: sebastien.guenneau@cnrs.fr}}
\affil{\textsuperscript{a}CNRS, Aix-Marseille Universit\'e, Centrale Marseille, Institut Fresnel, 13013 Marseille, France; \\
\textsuperscript{b}School of Aerospace Science and Technology, Xidian University, Xi’an 710071, China; \\
\textsuperscript{c}UMI 2004 Abraham de Moivre-CNRS, Imperial College London, SW7 2AZ, United Kingdom}
}

\maketitle

\begin{abstract}
We extend results of [Liu et al. Proc. Roy. Soc. Lond. A 469, 20130240, 2013] on artificial magneto-electric coupling and magnetism in moderate contrast dielectric layered media via high-order homogenization to the three-dimensional setting. For this, we consider asymptotic expansions in the vector Maxwell system. We then illustrate effect of magneto-electric coupling with numerical simulations for a layered system alternating positively and negatively refracting index. We unveil hyperbolic behaviour and magneto-electric coupling and derive a sharper estimate for the anisotropic effective medium in [Ramakrishna et al., J. Mod. Optics, 50, 1419-1430, 2003]. We also revisit the effective medium description of a PT-symmetric layered structure in [Novitsky, Shalin, Novitsky, Phys. Rev. A 99, 043812, 2019], adding higher-order approximation orders.  

\end{abstract}


\section{Introduction}

There is currently renewed interest in artificial magnetism in high-contrast photonic crystals, and metamaterials, notably in the quasi-static regime. Artificial bianisotropy is another very active research area of effective medium theories in periodic structures, such as arrays of split ring resonators or swiss-rolls. However, it is not widely appreciated that there is even magnetic activity and magneto-electric coupling in low contrast layered media \cite{liu2013a}.

A readily accessible asymptotic regime through effective medium theories, much studied in the literature on multipole expansions, is for low frequency (quasi-static), or dilute composite, limits. Alternatively if one introduces a small positive parameter $\eta$, say a ratio of the array pitch by the wave wavelength, then so called two-scale asymptotic expansions can be used to identify leading order homogenized Maxwell's equations, and in this way, effective properties are obtained at fixed frequency \cite{guenneau2000,wellander2001homogenization}, that involve an artificially anisotropic effective permittivity. In case of high-contrast in the material parameters, say when a highly-conducting inclusion is surrounded by a dielectric matrix, two-scale expansions unveil artificial magnetism \cite{felbacq2005}. An alternative route to achieve not only artificial magnetism, but also magneto-electric coupling, is via high order homogenization as proposed in \cite{liu2013a} with a homogenization algorithm using the transfer matrix formalism. This high-order homogenization algorithm has been used subsequently in \cite{liu2013PRB,popov2016operator,Novitsky2019PT,lebbe2022,cornaggia2022homogenized}.  However, the latter route becomes cumbersome with doubly and triply periodic structures. We thus revert to two-scale expansions, however with higher order corrections taken into account in the homogenized Maxwell system, as proposed for the conductivity equation in \cite{bakhvalov89}.

\section{High-order homogenization of constitutive equations}

We consider a finite photonic crystal (PC) that consists of a large number of periodic cells $\eta Y={[0,\eta]}^3$, filled with a dielectric medium with permittivity $\varepsilon({\bf x}/\eta)$, which is periodic of period $\eta$ in the bounded domain $\Omega$ (the PC) and which is equal to $\varepsilon_0$ in the unbounded domain outside $\Omega$. As per usual one denotes by $\varepsilon_0\mu_0=c^{-2}$, where $c$ is the velocity of light in vacuum. Since we consider a dielectric, non-magnetic medium, illuminated by a time harmonic wave, the oscillatory electromagnetic field within the PC $({\bf E}_\eta,{\bf H}_\eta)$ satisfies the first two Maxwell's equations
\begin{equation}
\begin{array}{l}
\nabla_{\bf x} \times {\bf E}_\eta({\bf x}) = i\omega\mu_0 {\bf H}_\eta({\bf x}) \, , \\ 
\nabla_{\bf x} \times {\bf H}_\eta({\bf x}) = - i\omega\varepsilon({\bf x}/\eta) {\bf E}_\eta({\bf x}) \, , 
\end{array}
\label{maxwellequations}
\end{equation}
where $\omega$ is the wave angular frequency (rad/s).
Taking the divergence of (\ref{maxwellequations}), we can proceed with the {\bg{following conditions}}
\begin{equation}
\nabla_{\bf x} \cdot \big[ \mu_0 {\bf H}_\eta({\bf x})\big] =0\quad {\rm and} \quad
\nabla_{\bf x} \cdot \big[ \varepsilon({\bf x}/\eta){\bf E}_\eta({\bf x})\big] =0
\label{divMaxwell}
\end{equation}
We assume the same asymptotic expansions of the {\bg{electric and magnetic oscillatory fields ${\bf E}_\eta$ and ${\bf H}_\eta$ as in \cite{guenneau2000}
\begin{equation}
{\bf E}_\eta({\bf x})={\bf E}_0({\bf x},{\bf x}/\eta)+\sum_{i=1}^N\eta^i {\bf E}_i({\bf x},{\bf x}/\eta) + O(\eta^N)
\label{ansatzE}
\end{equation}
and 
\begin{equation}
{\bf H}_\eta({\bf x})={\bf H}_0({\bf x},{\bf x}/\eta)+\sum_{i=1}^N\eta^i {\bf H}_i({\bf x},{\bf x}/\eta) + O(\eta^N)
\label{ansatzH}
\end{equation}
One notes that ${\bf E} _i$ and ${\bf H} _i$}} are periodic of period $\eta$ in the second variable and besides from that {\bg{${\bf E} _\eta$ and ${\bf H} _\eta$}} should be of finite energy on every compact set, and satisfy some outgoing wave condition, so that the diffraction problem of a plane wave illuminating the PC be well posed (existence and uniqueness of the solution) for each $\eta$. We refer to \cite{guenneau2000,wellander2001homogenization} for such questions.

Let us adopt the following notations:
\begin{equation}
{F}_\eta({\bf x}) = \left[
\begin{array}{l}
{\bf E}_\eta({\bf x}) \\ 
{\bf H}_\eta({\bf x}) 
\end{array}\right] 
= {F}_0({\bf x}, {\bf x}/\eta) 
+ \sum_{i=1}^N\eta^i F_i({\bf x},{\bf x}/\eta) + O(\eta^N)
\label{def-F}
\end{equation}
with
\begin{equation}
{F}_i({\bf x}, {\bf x}/\eta) = \left[
\begin{array}{l}
{\bf E}_i({\bf x}, {\bf x}/\eta) \\ 
{\bf H}_i({\bf x}, {\bf x}/\eta) 
\end{array}\right] \quad \text{for} \quad 
i = 0,1, \dots, N \, ,
\label{def-Fi}
\end{equation}
and 
\begin{equation}
M_{\bf x} = \left[
\begin{array}{ll}
0 & i \, \varepsilon_0^{-1} \nabla_{\bf x} \times \\
- i \, \mu_0^{-1} \nabla_{\bf x} \times & 0 
\end{array} \right] \, , \quad
V({\bf x}/\eta) = - \omega \dfrac{\varepsilon({\bf x}/\eta) - \varepsilon_0}{\varepsilon_0} \left[
\begin{array}{ll}
1 & 0 \\
0 & 0 
\end{array} \right] \, .
\label{def-MV}
\end{equation}
Then, the Maxwell's equations (\ref{maxwellequations}) can be 
written as
\begin{equation}
 \omega F_\eta({\bf x}) = M_{\bf x} \, F_\eta({\bf x})
+ V({\bf x}/\eta) \, F_\eta({\bf x}) \, .
\label{ME-FMV}
\end{equation}
Rescaling the differential operator $M_{\bf x}$ as $M_{\bf x}+\eta^{-1}M_{\bf y}$ and plugging the asymptotic {\bg{expansion (\ref{def-F}) in the Maxwell's equations (\ref{ME-FMV}) and further collecting terms of factors of same powers of $\eta$, we get a hierarchy of equations {\bg{in powers of $\eta$ with, at the leading order in 
$\eta^{-1}$,}}
\begin{equation}
\begin{array}{ll}
M_{\bf y} \, F_0({\bf x}, {\bf y}) = 0 
\end{array}
\label{ME-order0}
\end{equation}
and, at the next orders,
\begin{equation}
\begin{array}{ll}
\eta^{0} & \hspace{-2mm}: \: \omega F_0({\bf x}, {\bf y}) =
M_{\bf x} F_0({\bf x}, {\bf y}) + 
M_{\bf y} F_1({\bf x}, {\bf y}) + 
V({\bf y}) F_0({\bf x}, {\bf y}) \, , \\[2mm]
\eta^{1} & \hspace{-2mm}: \: \omega F_1({\bf x}, {\bf y}) =
M_{\bf x} F_1({\bf x}, {\bf y}) + 
M_{\bf y} F_2({\bf x}, {\bf y}) + 
V({\bf y}) F_1({\bf x}, {\bf y}) \, , \\[2mm] 
\eta^{2} & \hspace{-2mm}: \: \omega F_2({\bf x}, {\bf y}) =
M_{\bf x} F_2({\bf x}, {\bf y}) + 
M_{\bf y} F_3({\bf x}, {\bf y}) + 
V({\bf y}) F_2({\bf x}, {\bf y}) \, , \\[2mm]
& \dots  \\[2mm]
\eta^{i} & \hspace{-2mm}: \: \omega F_i({\bf x}, {\bf y}) =
M_{\bf x} F_i({\bf x}, {\bf y}) + 
M_{\bf y} F_{i+1}({\bf x}, {\bf y}) + 
V({\bf y}) F_i({\bf x}, {\bf y}) \, .
\end{array}
\label{ME-orders}
\end{equation}
Applying the mean operator $\bla \cdot \bra$ over the periodic cell $Y={[0,1]}^3$, and invoking the generalized Stokes' theorem applied over $Y$ (a torus, so a manifold without boundary), this set of equations (\ref{ME-orders}) above imply
\begin{equation}
\begin{array}{ll}
\eta^{0} & \hspace{-2mm}: \: 
\omega \, \bla F_0 ({\bf x}, {\bf y}) \bra =
M_{\bf x} \bla F_0({\bf x}, {\bf y}) \bra + 
\bla V({\bf y}) F_0({\bf x}, {\bf y}) \bra \, , \\[2mm]
\eta^{1} & \hspace{-2mm}: \: 
\omega \, \bla F_1 ({\bf x}, {\bf y}) \bra =
M_{\bf x} \bla F_1({\bf x}, {\bf y}) \bra + 
\bla V({\bf y}) F_1({\bf x}, {\bf y}) \bra  \, , \\[2mm] 
\eta^{2} & \hspace{-2mm}: \: 
\omega \, \bla F_2 ({\bf x}, {\bf y}) \bra =
M_{\bf x} \bla F_2({\bf x}, {\bf y}) \bra + 
\bla V({\bf y}) F_2({\bf x}, {\bf y}) \bra  \, , \\[2mm]
& \dots  \\[2mm]
\eta^{i} & \hspace{-2mm}: \: 
\omega \, \bla F_i ({\bf x}, {\bf y}) \bra =
M_{\bf x} \bla F_i({\bf x}, {\bf y}) \bra + 
\bla V({\bf y}) F_i({\bf x}, {\bf y}) \bra  \, .
\end{array}
\label{ME-hom}
\end{equation}
These last equations provide the different orders of the 
homogenized Maxwell's equations. 

Before applying this mean operator $\bla \cdot \bra$ over 
the periodic cell $Y={[0,1]}^3$, the homogenization procedure 
requires to derive the expression of the components 
$F_i({\bf x}, {\bf y})$ of $F_\eta({\bf x}, {\bf x}/\eta)$. 
For the leading order, this derivation starts with 
the equations (\ref{ME-order0}) and (\ref{ME-orders}) 
for the transverse part and, for the longitudinal 
part, with the equations deduced from 
(\ref{divMaxwell}), see the sections 5.1 in the annex for details. 

For the leading order, the equation (\ref{ME-order0}) 
and the equation deduced from (\ref{divMaxwell}) 
show that the electric and magnetic components 
${\bf E}_0({\bf x}, {\bf y})$ and 
${\bf H}_0({\bf x}, {\bf y})$ are uncoupled. 
However, it is stressed that, for the next orders, the equations 
(\ref{ME-orders}) shows that the transverse parts 
of the electric and magnetic components 
${\bf E}_i({\bf x}, {\bf y})$ and 
${\bf H}_i({\bf x}, {\bf y})$ are generated by source-terms ${\bf H}_{i-1}({\bf x}, {\bf y})$ and 
${\bf E}_{i-1}({\bf x}, {\bf y})$. Therefore, we have
formally that 
\begin{equation}
\begin{array}{ll}
{\bf E}_i = W_i^{ee} {\bf E}_{i-1} + W_i^{eh} {\bf H}_{i-1} \, , \\[2mm]
{\bf H}_i = W_i^{he} {\bf E}_{i-1} + W_i^{hh} {\bf H}_{i-1} \, . \\
\end{array}
\label{formal}
\end{equation}
According to the homogenized Maxwell's equations (\ref{ME-hom}), the effective parameters describing the interaction between the electromagnetic field and the media are given at each order by
\begin{equation}
\bla V({\bf y}) F_i({\bf x}, {\bf y}) \bra \, , 
\label{const-i}
\end{equation}
and the different orders of the  homogenized electric displacement 
field and magnetic inductance are given by:
\begin{equation}
\bla \big[ \omega - V({\bf y}) \big] F_i({\bf x}, {\bf y}) \bra \, . 
\label{const-i2}
\end{equation}
Hence, from the formal expressions (\ref{formal}), it appears that, 
the equation above is reminiscent of the structure of a constitutive equation for an effective bianisotropic medium as in {\color{black}{\cite{kong86,tretyakov,liu2013a}}}\begin{equation}
\begin{array}{l}
{\bf D}_{hom}({\bf x}) =\varepsilon_{hom} {\bf E}_{hom}({\bf x}) + {\color{black}{ \Xi_{hom} }} {\bf H}_{hom}({\bf x}) \, , \\[2mm]
{\bf B}_{hom}({\bf x}) =\mu_{hom} {\bf H}_{hom}({\bf x}) + {\color{black}{ ^{t}\Xi_{hom} }} {\bf E}_{hom}({\bf x}) \, ,
\end{array}
\label{hombianisotropic} 
\end{equation}
where ${\bf D}_{hom}$ and ${\bf B}_{hom}$ are homogenized electric displacement field and magnetic inductance, $\Xi_{hom}$ and its symmetric $^{t}\Xi_{hom}$ are rank-2 tensors of magnetic-electric coupling. Notice that, at the first order, the magnetic field 
${\bf H}_1({\bf x},{\bf y})$ is the solution to this equations for 
its transverse and longitudinal parts \cite{GGT07}
\begin{equation}
\nabla_{\bf y} \times \nabla_{\bf y} \times 
{\bf H}_1({\bf x},{\bf y})=-i\omega 
\nabla_{\bf y} \times \varepsilon({\bf y}) {\bf E}_0({\bf x},{\bf y}) \, , \quad \quad 
\nabla_{\bf y} \cdot  {\bf H}_1({\bf x},{\bf y})=0
\label{eq-H1} 
\end{equation}
since ${\bf H}_0({\bf x},{\bf y}) = {\bf H}_0({\bf x}) = 
\bla {\bf H}_0 \bra ({\bf x})$ (see section 5.2 in the annex). The 
component ${\bf H}_1({\bf x},{\bf y})$ is thus generated by the 
solely source term ${\bf E}_0({\bf x},{\bf y})$, which means that 
the effective constitutive equation for ${\bf B}_{hom}({\bf x})$ 
reduces to a term driven by ${\bf E}_{hom}({\bf x})$. As a 
consequence, $\mu_{hom}$ remains the vacuum permeability $\mu_0$ 
at the order 1, and the effective permeability $\mu_{hom}$ differs 
from $\mu_0$ only from the order 2. This is in agreement with 
the results stated in \cite{liu2013a}, where the effective 
magneto-electric coupling is obtained from order 1 and then 
the effective magnetism from order 2. 

\section{Illustrative application : the layered case}

Derivation of the homogenized parameters in the layered case is quite straightforward, a detailed derivation can be found in \cite{liu2013a, liu2013PRB}. The interested reader can find details on how to compute ${\bf V}${\color{black}{$^{(0)}$}} in (\ref{leadingannex}), a complete derivation of the leading order expression for the homogenized permittivity (\ref{leadingpermittivity}) in \cite{pavliotis2010}, and for higher-order expressions we refer to \cite{allaire2018optimization}. 

\begin{figure}[!h]
\centering
\includegraphics[scale=0.8]{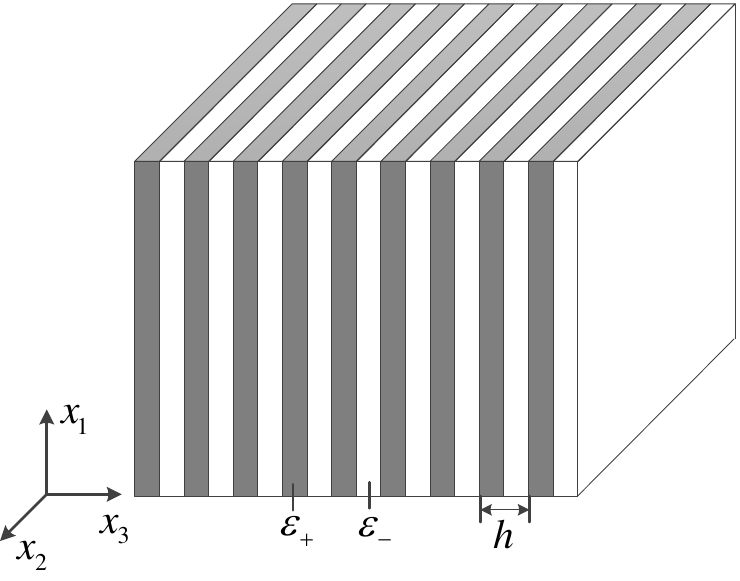}
\caption{Schematic diagram of a layered Poor Man's lens, the permittivities of two dielectrics are $\varepsilon_+$ and $\varepsilon_-$, the thicknesses of two layers are the same.}
\label{poorlen}
\end{figure}

\subsection{Hyperbolicity and optical activity in the Poor Man's lens}

We apply the above HOH to a layered Poor Man's lens \cite{Pendry2000}, shown in Fig.\ref{poorlen}, which consists of an alternation of two dielectrics with permittivity equal to $\varepsilon_1/\varepsilon_0(=+1)$ and $\varepsilon_2/\varepsilon_0(=-1)$, and filling fraction are $f_1(=1/2)$ and $f_2(=1/2)$ respectively.

The zeroth order approximation provides us with the same expressions as the classical homogenization
\begin{equation}
\begin{array}{l}
\varepsilon_\parallel= \varepsilon_\perp = \varepsilon_1 f_1 + \varepsilon_2 f_2=0\,, \\[2mm]
\varepsilon_3^{-1}=\varepsilon^{-1}_1f_1 + \varepsilon^{-1}_2f_2=0 \,, \\[2mm]
\mu_\parallel = \mu_\perp = \mu_3 = \mu_0\,.
\end{array}
\label{eff0}
\end{equation}
If we move forward to the first order approximation, the permittivity and permeability are the same as in Eq.(\ref{eff0}), while we obtain the magneto-electric coupling as follows
\begin{equation}
\begin{array}{l}
K_\perp= \displaystyle\frac{\omega h}{2}\mu_0(\varepsilon_1-\varepsilon_2) f_1f_2 =\displaystyle\frac{\omega h}{4c^2}\,, \\[3mm]
K_\parallel= \displaystyle\frac{\omega h}{2} \mu_0(\varepsilon_1-\varepsilon_2) f_1f_2 \left[1-\displaystyle\frac{{k}^2}{{ \omega}^2}\displaystyle\frac{\varepsilon_1+\varepsilon_2}{\mu_0\varepsilon_1\varepsilon_2}\right]=\displaystyle\frac{\omega h}{4c^2}\,.
\end{array}
\label{eff1}
\end{equation}
where ${k}^2={\boldsymbol k}\cdot {\boldsymbol k}$ and ${\boldsymbol k}$ is the two-dimensional component of the three-dimensional wave vector ${\boldsymbol{k}}$ after projection in the plane $(x_1,x_2)$. Note that 
the magnetoelectric coupling parameter is isotropic at this order, since $K_\perp= K_\parallel$ in this very specific sign-shifting configuration where $\varepsilon_1 = - \varepsilon_2$.
Further, considering the second order approximation, the effective permittivity and permeability expressions
are \cite{liu2013a}
\begin{equation}
\begin{array}{ll}
\varepsilon_\parallel \hspace*{-5mm}&= \varepsilon_1 f_1 + \varepsilon_2 f_2 + \displaystyle\frac{\omega^2 h^2 }{6} \mu_0 f_1 f_2 (\varepsilon_1-\varepsilon_2) (\varepsilon_1 f_1-\varepsilon_2 f_2) 
\left(1-\displaystyle\frac{{ k}^2}{{\omega}^2} 
\displaystyle\frac{\varepsilon_1+\varepsilon_2}{\mu_0\varepsilon_1 \varepsilon_2}\right) = \displaystyle\frac{\omega^2h^2}{12c^2}\,\varepsilon_0 \, ,\\[3mm]
\varepsilon_\perp \hspace*{-5mm} & = \varepsilon_1 f_1 + \varepsilon_2 f_2 + \displaystyle\frac{\omega^2 h^2}{6} \mu_0 f_1 f_2 (\varepsilon_1-\varepsilon_2) (\varepsilon_1 f_1-\varepsilon_2 f_2) = \displaystyle\frac{\omega^2h^2}{12c^2}\, \varepsilon_0 \, ,\\[3mm]
{\varepsilon_3}^{-1} \hspace*{-3mm} & = {\varepsilon^{-1}_1}{f_1} +{\varepsilon^{-1}_2}{f_2} - \displaystyle\frac{\omega^2 h^2}{6} \mu_0 f_1 f_2 (\varepsilon_1-\varepsilon_2) (\displaystyle\frac{f_1}{\varepsilon_1}  -\displaystyle\frac{f_2}{\varepsilon_2}) \left(1-\displaystyle\frac{{ k}^2}{{\omega}^2} \displaystyle\frac{\varepsilon_1+\varepsilon_2}{\mu_0\varepsilon_1 \varepsilon_2}\right) = - \displaystyle\frac{\omega^2h^2}{12c^2}\, 
\varepsilon_0 \, ,\\[3mm]
\mu_\parallel \hspace*{-5mm}& = \mu_0 - \displaystyle\frac{\omega^2 h^2}{6} \mu^2_0 f_1 f_2 (\varepsilon_1-\varepsilon_2) (f_1-f_2)=\mu_0 \,,\\[3mm]
\mu_\perp \hspace*{-5mm}& = \mu_0 - \displaystyle\frac{\omega^2 h^2}{6} \mu^2_0 f_1 f_2 (\varepsilon_1-\varepsilon_2) (f_1-f_2) \left(1-\displaystyle\frac{{k}^2}{{\omega}^2} \displaystyle\frac{\varepsilon_1+\varepsilon_2}{\mu_0\varepsilon_1 \varepsilon_2}\right) =\mu_0 \,, \\[3mm]
{\mu_3}^{-1} \hspace*{-3mm}& = {\mu^{-1}_0} + \displaystyle\frac{\omega^2 h^2}{6} f_1f_2(\varepsilon_1-\varepsilon_2) (f_1-f_2) = {\mu^{-1}_0}\,.
\end{array}
\label{eff2}
\end{equation}
We note that our effective medium is highly anisotropic when $\omega$ tends to $0$ but, unlike for the static limit studied in \cite{ramakrishna2003}, it is also hyperbolic since the $\varepsilon_3$ component takes negative values. This hyperbolic behaviour can be supported by the following argument. Let $\varepsilon_1 = \varepsilon_0$ and $\varepsilon_2 = - \varepsilon_0 (1- \delta)$, with $0<\delta\ll 1$. Then it is straightforward to check that, at the zeroth order, 
\begin{equation}
\begin{array}{l}
\varepsilon_\parallel= \varepsilon_\perp = \varepsilon_1 f_1 + \varepsilon_2 f_2=
\varepsilon_0 f_1 - \varepsilon_0 (1-\delta) f_2= \varepsilon_0 \delta f_2 \, , \\[2mm]
\varepsilon_3^{-1}=\varepsilon^{-1}_0f_1 - \varepsilon^{-1}_0 (1-\delta)^{-1}f_2\approx
- \varepsilon^{-1}_0 \delta f_2 \, , 
\end{array}
\label{eff0delta}
\end{equation}
which shows that both perturbative corrections, with respect to small frequencies (\ref{eff2}) 
and to small deviation from the sign shifting (\ref{eff0delta}), lead to a hyperbolic behaviour 
of the multilayered Poor Man's lens. 
Moreover, our modelling is different at the first-order correction (\ref{eff1}) since it 
brings some magnetoelectric coupling that is not accounted for in the effective medium model derived in \cite{ramakrishna2003}. This first order correction is not surprising since 
it has been known for over a decade that truncation of photonic crystals matters in their effective properties \cite{pierre2008appropriate,smigaj2008validity,simovski2010electromagnetic,yang2010retrieving,vinogradov2011additional}. Finally, we also notice that there is no effective magnetism, 
just like in \cite{ramakrishna2003}.

As a numerical illustration performed with Matlab, let us consider again the multilayered Poor Man's len, the total number of the unit layers is $N=20$; the parameters of the two layers constituting each unit layer are $\varepsilon_1/\varepsilon_0=1$, $\mu_1/\mu_0=1$, $\varepsilon_2/\varepsilon_0=-1+0.0001*i$, $\mu_2/\mu_0=1$, a small absorption being introduced to ensure the convergence in Matlab. Note that the p- and s-polarized incident waves coincide under a normal incidence ($\boldsymbol{k}={\bf 0}$), i.e. $t_{\rm p}=t_{\rm s}=t$. Now we would like to compare the transmission properties of the stack and its effective medium obtained by the above homogenization method. The transfer matrix method in \cite{liu2013a} is introduced to compute the transmission coefficients as a function of the normalized frequency $\omega h / (2 \pi c)\equiv \eta$, Fig. \ref{figtes} shows both the transmission curve of the stack (solid line) and that of the effective medium (dashed line) at the 2nd order approximation with parameters in Eqs.(\ref{eff0})-(\ref{eff2}). The ordinate in the figures is the real part of the transmission coefficient, while the abscissa is the normalized frequency $\eta$. We stress that, in this section, the non-dimensional parameter $\eta$ is the normalized frequency 
$\omega h / (2 \pi c)$, and it is thus 
different than the parameter $\eta$ used in the 
section 2 (that was related to a non-dimensional length); however, in both cases, the homogenization is studied for vanishing parameter $\eta$. 

It can be observed in Fig. \ref{figtes} that the transmission curve of the effective homogeneous medium at the second order approximation (dashed line) fits well with the curve of the stack (solid line) at the normalized frequencies $\eta$ lower than 0.4. With a higher order approximation, an improved agreement between the transmission curves of the stack and its effective medium can be achieved, see \cite{liu2013a}.
\begin{figure}[!htp]
\centering
\includegraphics[scale=0.4]{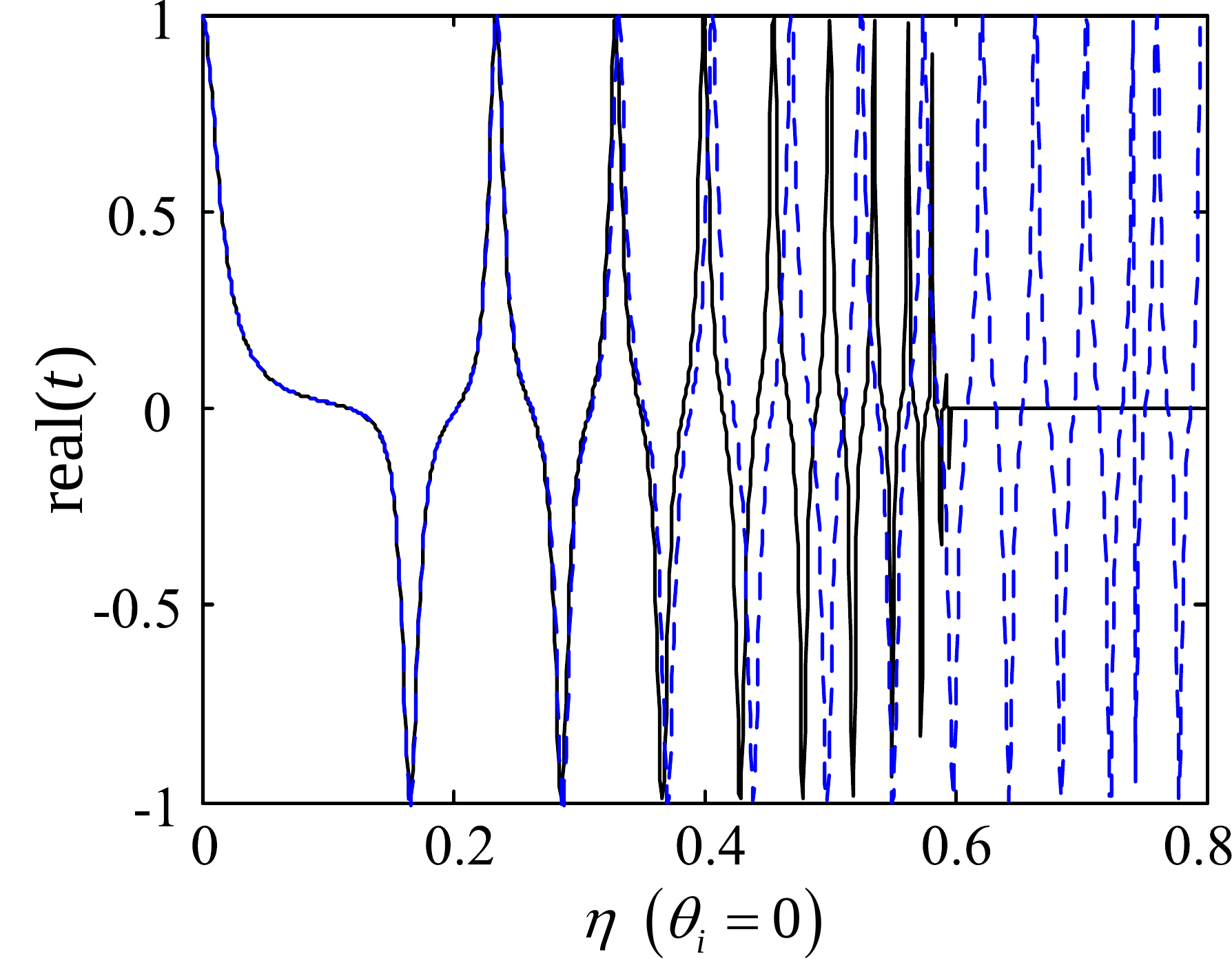}
\caption{Transmission curves of the multilayered stack (solid line) and effective medium at 2nd order (dashed line) approximation.The permittivities of the two dielectric layers are $\varepsilon_1/\varepsilon_0=1$, $\varepsilon_2/\varepsilon_0=-1+0.0001*i$, respectively.}
\label{figtes}
\end{figure}

Furthermore, considering an oblique incidence with $\theta_i=30^\circ$ in p-polarization, as well as s-polarization, the transmission curves of the stack and its effective medium are depicted in Fig. \ref{figtem}. The asymptotic approximation at the second order is in good agreement up to the normalized frequency 0.5 in s-polarization and 0.3 in p-polarisation. Notice that this numerical example confirms the hyperbolic feature of the Poor Man's lens at low frequencies.
\begin{figure}[htb]
\centering
\subfigure[{\color{black}{s-polarization}}]{
\includegraphics[width=7.0cm]{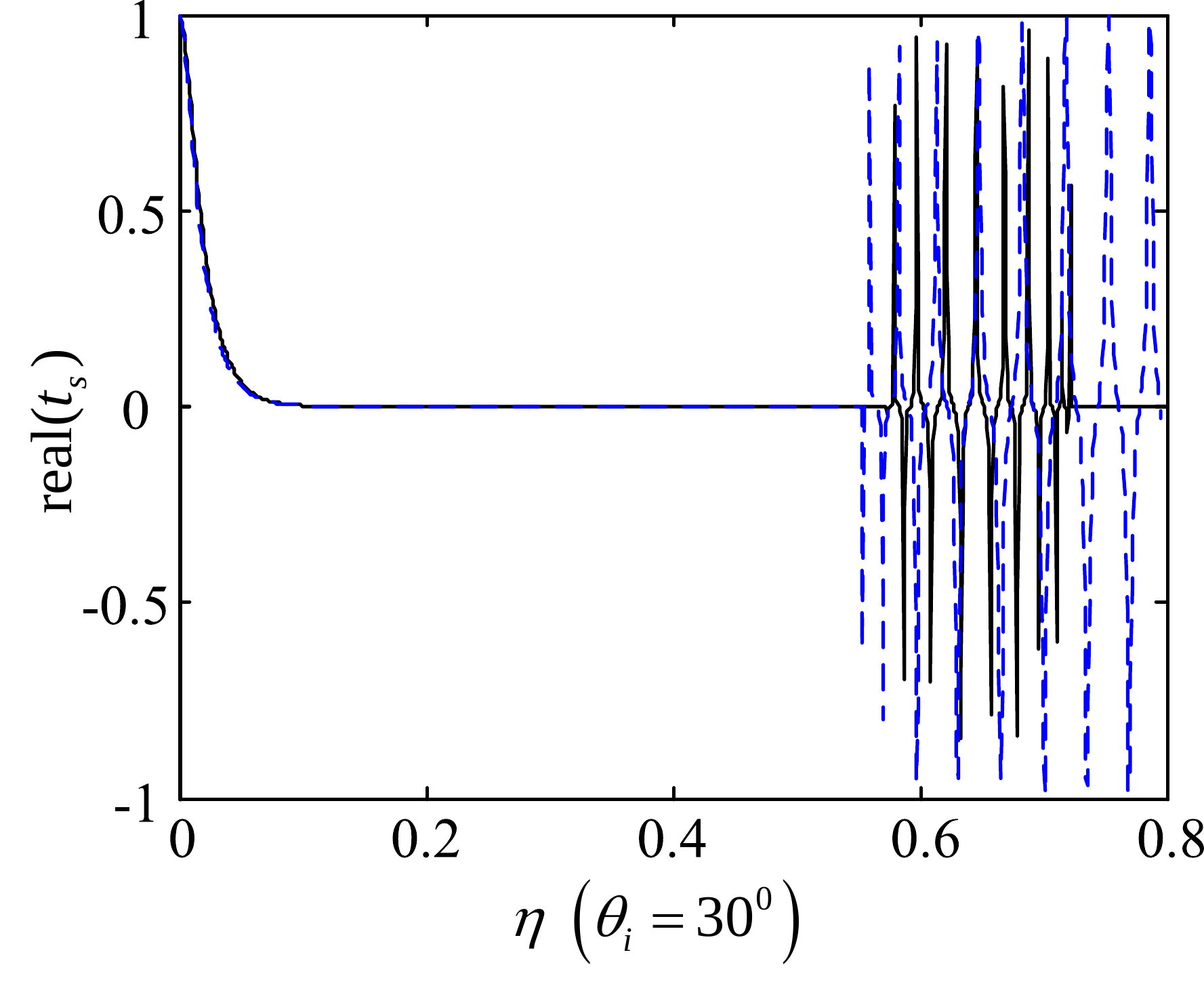}}  
\subfigure[{\color{black}{p-polarization}}]{
\includegraphics[width=6.8cm]{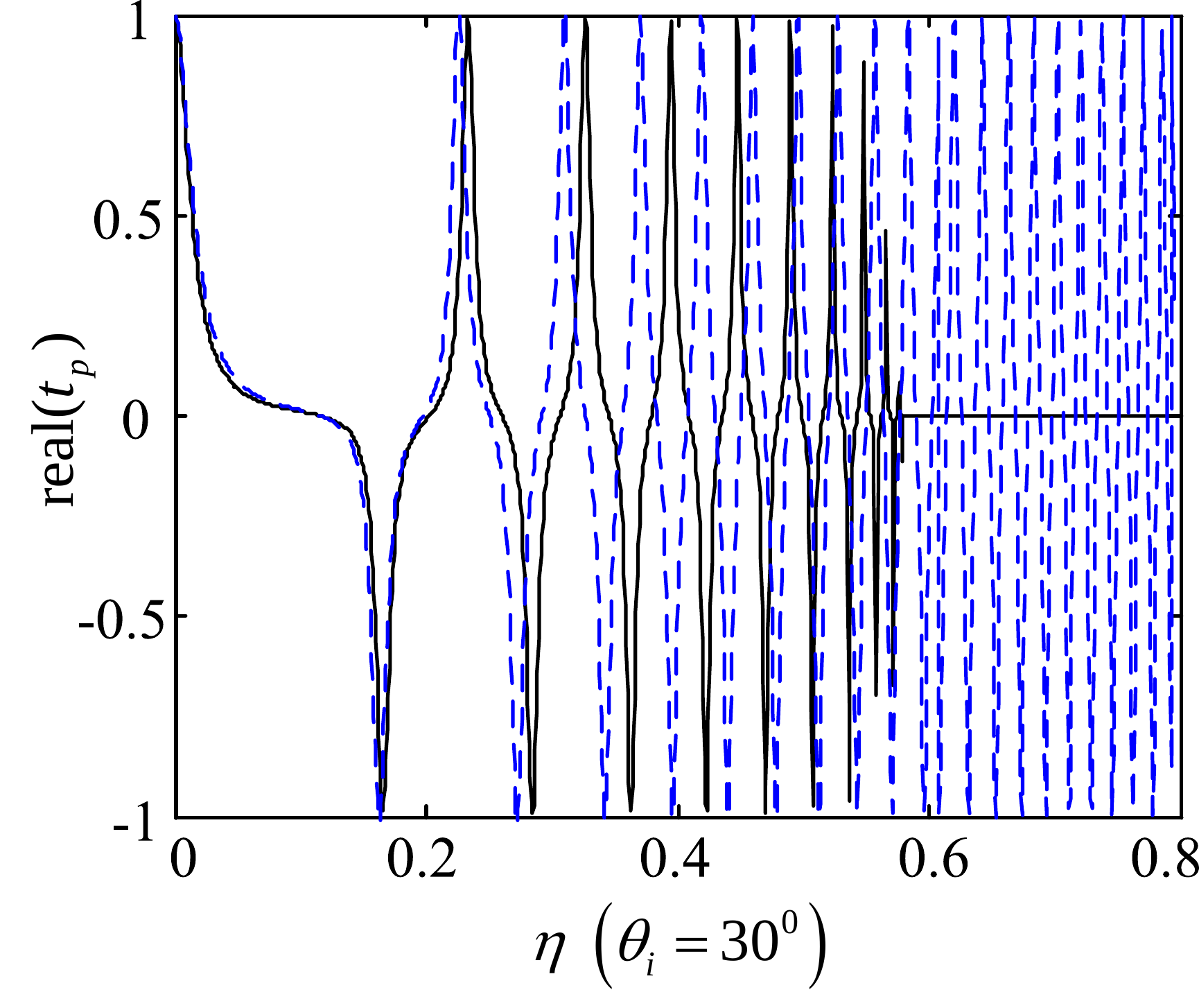}}
\caption{ Transmission curves of the multilayered stack (solid line) and effective medium at second order approximation (dashed line) as a function of the normalized frequency $\eta = \omega h / (2 \pi c)$ under an oblique incidence with $\theta_i=30^\circ$: (a) {\color{black}{s-polarization}}; (b) {\color{black}{p-polarization}}. The permittivities of the two layers are $\varepsilon_1/\varepsilon_0=1$, $\varepsilon_2/\varepsilon_0=-1+0.0001*i$, respectively.}
\label{figtem}
\end{figure}

\subsection{The P-T symmetric layered case}

We turn to a second application of the high order homogenization in the configuration the 
P-T symmetric system studied in \cite{Novitsky2019PT}. Fig.\ref{PTmultilayer} shows the considered P-T symmetric multilayered structure, which is a periodic structure with balanced loss and gain, i.e. the permittivities of the two component layers are $\varepsilon_{L}=\varepsilon'+i\varepsilon''$, $\varepsilon_{G}=\varepsilon'-i\varepsilon''$ with the same value of loss and gain coefficients. 
 \begin{figure}
\centering
\includegraphics[scale=0.5]{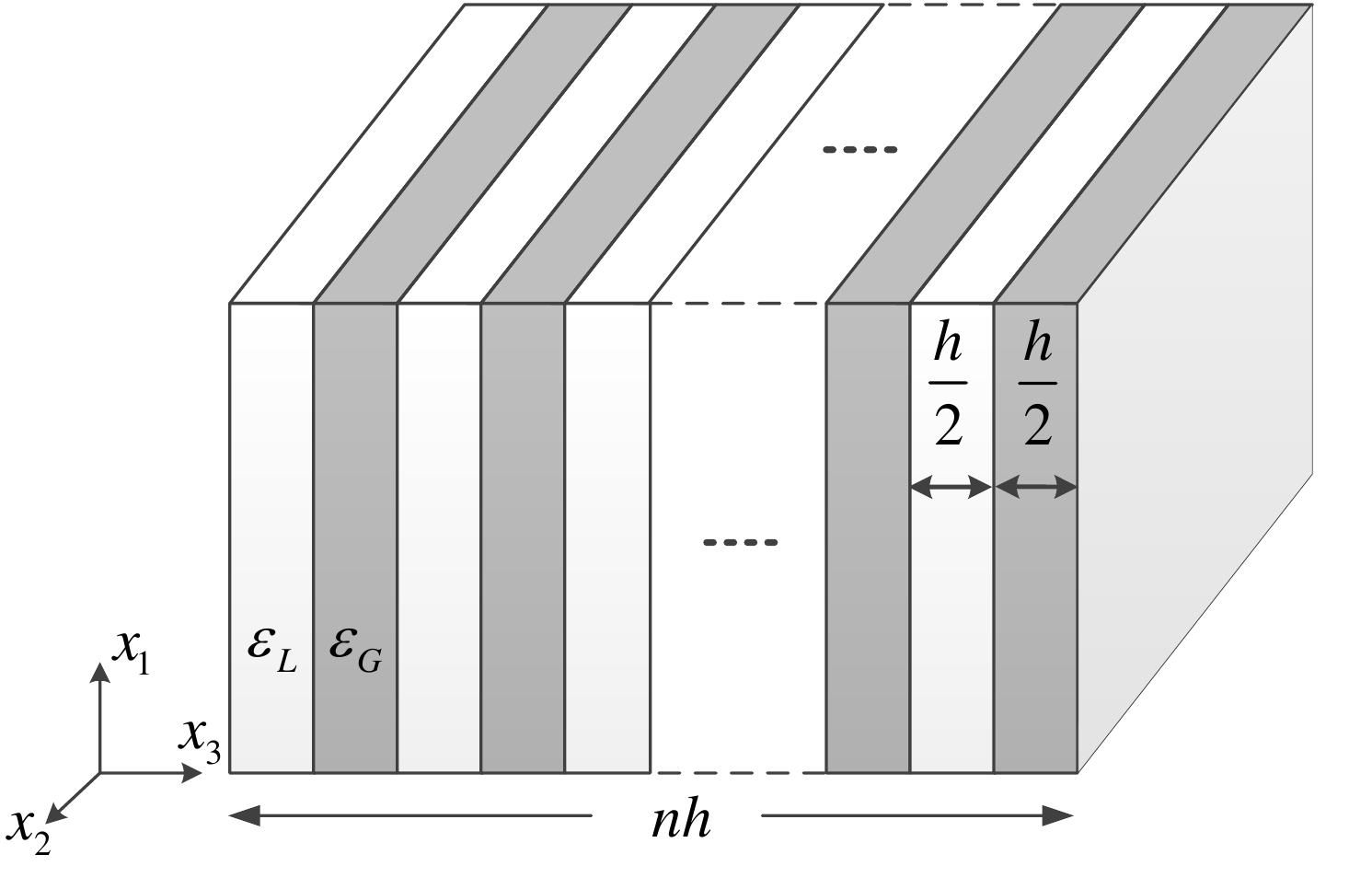}
\caption{Geometry of a finite P-T symmetric multilayered structure, the unit cell consists of two dielectrics with balanced loss and gain, i.e. $\varepsilon_{L}$ and $\varepsilon_{G}$, the thicknesses of two layers are the same, the total number of layers is 2N.}
\label{PTmultilayer}
\end{figure}
Same HOH method is applied to such a P-T multilayered structure, the expressions of effective parameters for classical approximation are
\beq
\begin{array}{l}
\varepsilon_\parallel= \varepsilon_\perp = \varepsilon_1 f_1 + \varepsilon_2 f_2=\varepsilon'\,, \\[2mm]
\varepsilon_3^{-1}=\varepsilon^{-1}_1f_1 + \varepsilon^{-1}_2f_2=\varepsilon' (\varepsilon'^2+\varepsilon''^2)^{-1} \,, \\[2mm]
\mu_\parallel = \mu_\perp = \mu_3 = \mu_0\,.
\end{array}
\label{PTeff0}
\eeq
while for the homogeneous medium at 1st order approximation, the entries of the magneto-electric coupling tensor are (in normal incidence)
\begin{equation}
\begin{array}{l}
K_\perp= K_\parallel= \displaystyle\frac{\omega h}{2} \mu_0(\varepsilon_1-\varepsilon_2) f_1f_2 = 
i \displaystyle\frac{\omega h}{4} \mu_0 \varepsilon'' \,. 
\end{array}
\label{PTeff1}
\end{equation}
When we consider the second order approximation, we obtain
\begin{equation}
\begin{array}{l}
\varepsilon_\parallel = \varepsilon_\perp= \varepsilon_1 f_1 + \varepsilon_2 f_2 + \displaystyle\frac{\omega^2 h^2 }{6} \mu_0 f_1 f_2 (\varepsilon_1-\varepsilon_2) (\varepsilon_1 f_1-\varepsilon_2 f_2) = \varepsilon' - \displaystyle\frac{\omega^2 h^2 }{12} \mu_0 
{\varepsilon''}^2
 \,,\\[3mm]
{\varepsilon_3}^{-1}= {\varepsilon^{-1}_1}{f_1} +{\varepsilon^{-1}_2}{f_2} - \displaystyle\frac{\omega^2 h^2}{6} \mu_0 f_1 f_2 (\varepsilon_1-\varepsilon_2) (\displaystyle\frac{f_1}{\varepsilon_1}  -\displaystyle\frac{f_2}{\varepsilon_2}) = \left( \varepsilon' - \displaystyle\frac{\omega^2 h^2 }{12} \mu_0 
{\varepsilon''}^2 \right) \displaystyle\frac{1}{\varepsilon'^2+\varepsilon''^2}
\,,\\[3mm]
\mu_\parallel = \mu_\perp= \mu_0 - \displaystyle\frac{\omega^2 h^2}{6} \mu^2_0 f_1 f_2 (\varepsilon_1-\varepsilon_2) (f_1-f_2) = \mu_0\,,\\[3mm]
{\mu_3}^{-1} = {\mu^{-1}_0} + \displaystyle\frac{\omega^2 h^2}{6} f_1f_2(\varepsilon_1-\varepsilon_2) (f_1-f_2) = {\mu^{-1}_0} \,.
\end{array}
\label{PTeff2}
\end{equation}
For the next orders, we refer to \cite{liu2013a}. 
\begin{figure}[htbp]
\centering
\subfigure[d=100nm]{
\includegraphics[width=6.8cm]{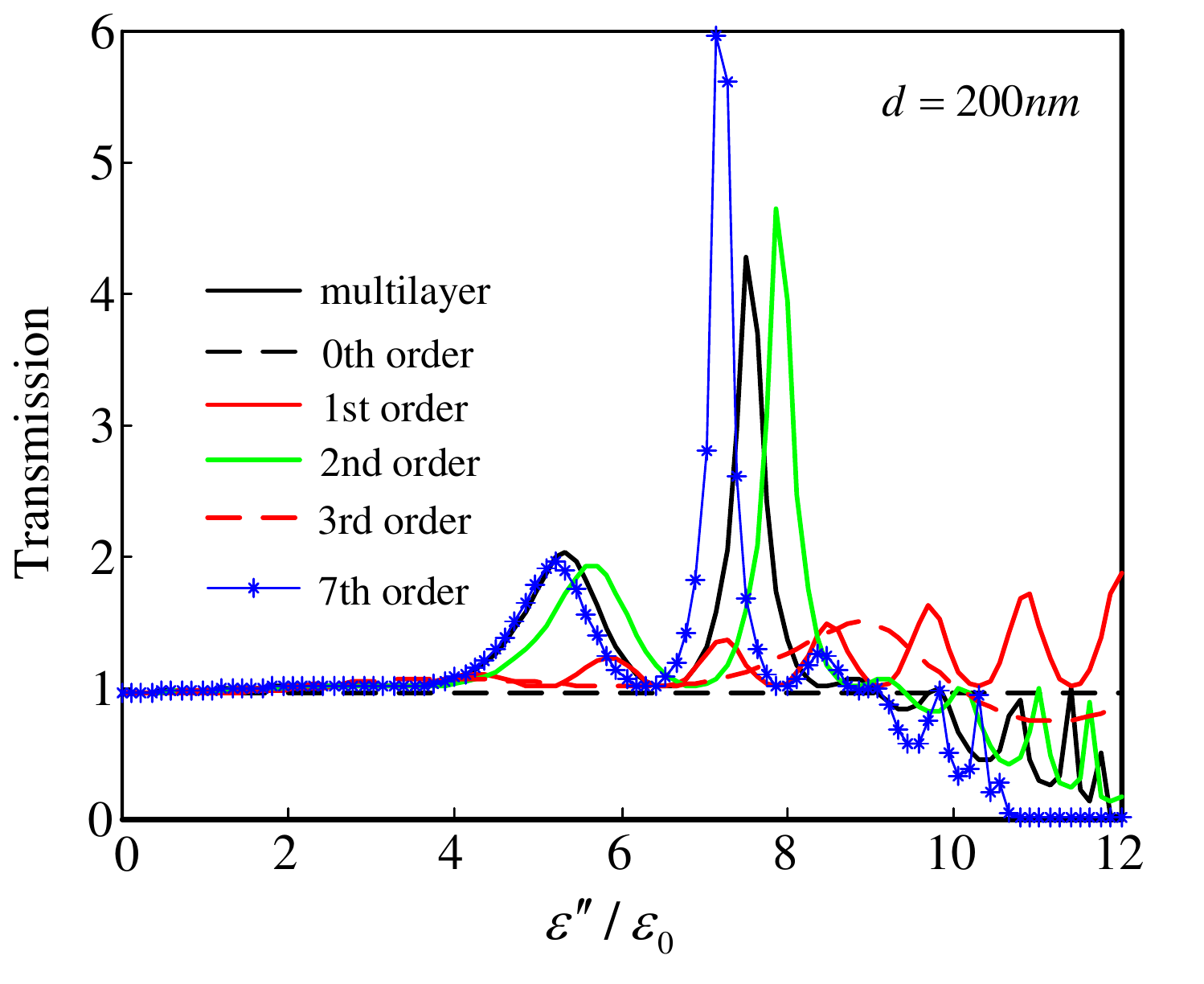}}
\subfigure[d=200nm]{
\includegraphics[width=6.8cm]{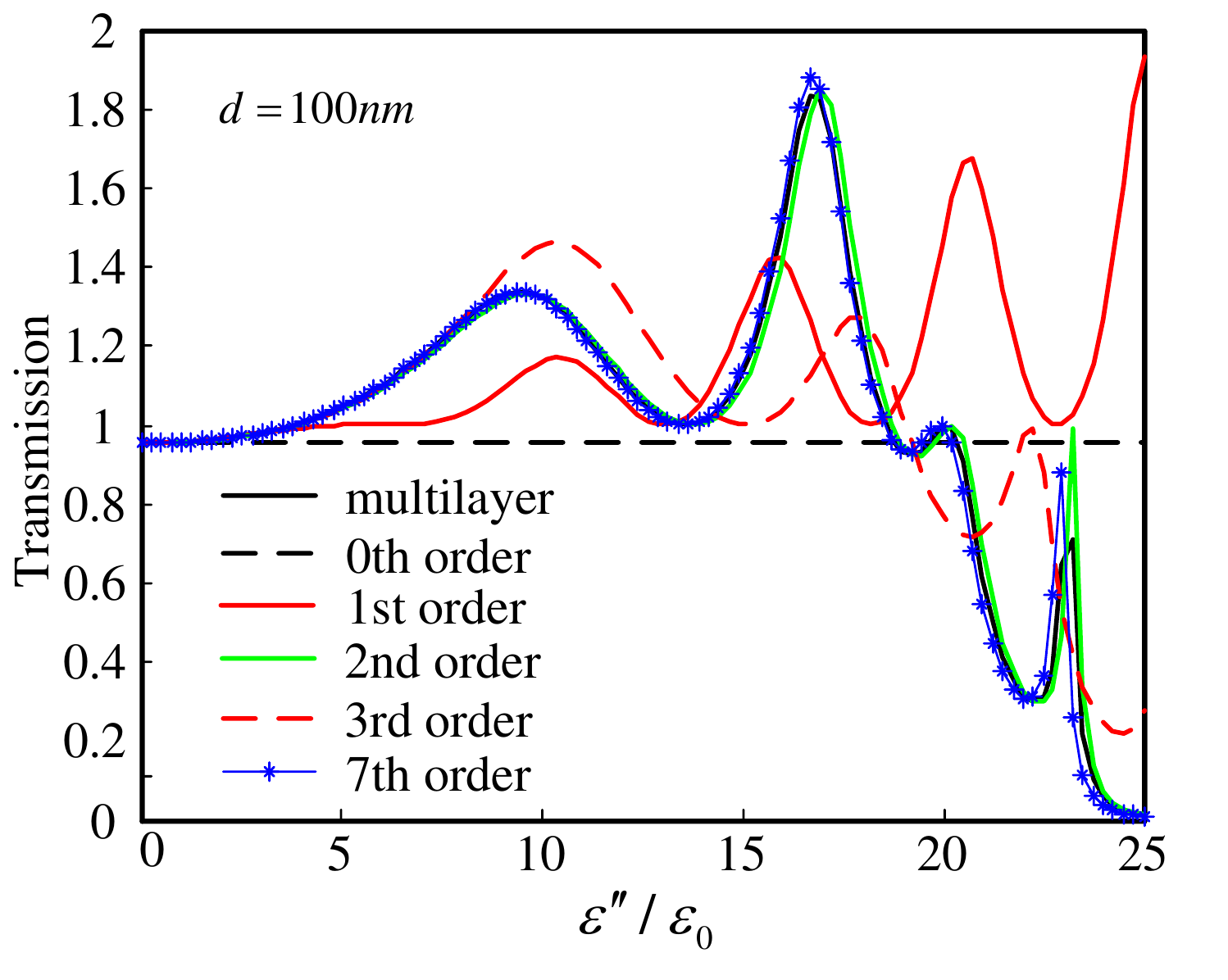}}
\caption{Transmission curves of PT-symmetric multilayer and the effective medium at 0th, 1st, 2nd, 3rd and 7th orders approximation as a function of $\varepsilon''/\varepsilon_0$. 
The permittivities of the two dielectric layers are $\varepsilon_1/\varepsilon_0=2+i\varepsilon''/\varepsilon_0$, $\varepsilon_2/\varepsilon_0=2-i\varepsilon''/\varepsilon_0$, respectively. 
The period thickness (a) d=100nm and (b) d=200nm. A normal incident wave with $\lambda=1.55\mu m$ is considered, the multilayered structure consists of N=20 slabs. }
\label{figPT}
\end{figure}

Fig. \ref{figPT} shows the comparison of transmission for the multilayer and for the effective medium up to the seventh order in the P-T symmetry configuration proposed in \cite{Novitsky2019PT}. 
The results are in good agreement and demonstrate the validity of the present high order homogenization in this P-T symmetry configuration. In particular, the seventh order bring a significative improvement of the approximation for values of $\varepsilon''/\varepsilon_0$ below 20 for $d=100$nm and below 6 for $d=200$nm.

\section{Conclusion}
In this paper, we introduced a high-order homogenization for the Maxwell's equations, devoted to constitutive equations of periodic dielectric media. We have shown that at leading order, one simply achieves some anisotropic permittivity. However, at the first order some magneto-electric coupling occurs as the homogenized constitutive equation has the structure of that for a bianisotropic medium. At the second order some artificial magnetism occurs. Interestingly, if one would consider some periodic medium with spatially varying permittivity and permeability, the proposed algorithm would be slightly modified as in (\ref{eq-order0}) the equations at order $\eta^{-1}$ would lead to a leading order asymptotic expansion of the form ${\bf H}_0({\bf x},{\bf y})=\bla {\bf H}_0 \bra ({\bf x}) + \nabla_{\bf y} [{\bf V}({\bf y})\cdot \bla {\bf H}_0\bra ({\bf x})]$, and the same annex problem as (\ref{leadingannex}) and some anisotropic permeability would be achieved at leading order as in (\ref{leadingpermittivity}). The homogenized Maxwell system at leading order was already derived for spatially varying permittivity and permeability using asymptotic expansions of the electric and magnetic fields (\ref{ansatzE})-(\ref{ansatzH}) at first order in all four Maxwell's equations (\ref{maxwellequations})-(\ref{divMaxwell}) \cite{wellander2001homogenization} and in the curlcurl Helmholtz operator by substituting the second equation into the first one in (\ref{maxwellequations}) thus working with the magnetic field unknown \cite{guenneau2007}. 
However, our derivation is based on higher-order terms in the asymptotic expansions of the Maxwell's equations (\ref{maxwellequations})-(\ref{divMaxwell}), and a series of ansatz at higher orders for the electric and magnetic field unknowns allows to identify higher order homogenized effective parameters.
If one proceeds with the high-order homogenization in the constitutive laws when both permittivity and permeability are heterogeneous, magneto-electric coupling occurs at first order as in the present case. We then recall the effective parameters for a layered structure at second order approximation already derived in \cite{liu2013a}, which has been reproduced with the exact same derivation in \cite{popov2016operator}. We retrieve the numerical results of 
\cite{Novitsky2019PT} for a P-T symmetric layered structure at second order approximation, and show improved results at seventh order. We also apply the high-order homogenization to a layered medium alternating layers of identical thickness with positive and negative permittivity (known as the layered Poor Man's lens) and at leading order we retrieve the effective medium permittivity in \cite{ramakrishna2003}. However, we find that some magneto-electric coupling occurs at the first order and, at second order, there is no magnetic activity. Thus the layered Poor Man's lens behaves effectively as a chiral medium with extreme effective permittivity, and moreover it is shown to be hyperbolic.

Our approach is different from that in \cite{maurel2018sensitivity} which considers some high-order homogenization for the electromagnetic problem based on a dynamic correction encapsulated in the boundary conditions at the interface between the homogenized stack and surrounding medium. Other works on improved effective medium approximation of layered media include \cite{GM12,popov2016operator,gorlach2020boundary,gralak2020,lannebere2020first}.
 

\bibliography{biblio.bib}



\section{Annex: Asymptotics in Maxwell's equations}

\subsection{Homogenized permittivity at leading order}

Rescaling the differential operator $\nabla_{\bf x}$ as $\nabla_{\bf x}+\eta^{-1}\nabla_{\bf y}$ and plugging the asymptotic {\bg{expansions (\ref{ansatzE}) and (\ref{ansatzH}) in the Maxwell's equations (\ref{maxwellequations}) and the resulting conditions (\ref{divMaxwell}),}} and further collecting terms of factors of same powers of $\eta$, we get a hierarchy of equations {\bg{in powers of $\eta$ with, at the leading order in 
$\eta^{-1}$,}}
\begin{equation}
\begin{array}{ll}
\nabla_{\bf y} \times  {\bf E}_0({\bf x},{\bf y})={\bf 0} \, , \\[2mm]
\nabla_{\bf y} \times  {\bf H}_0({\bf x},{\bf y})={\bf 0} \, , \\[2mm]
\nabla_{\bf y} \cdot \big[ \varepsilon({\bf y}) {\bf E}_0({\bf x},{\bf y}) \big] =0 \, , \\[2mm]
\nabla_{\bf y} \cdot  {\bf H}_0({\bf x},{\bf y})=0 \, ; \\
\end{array}
\label{eq-order0}
\end{equation}
{\bg{and, at the next order in 
$\eta^{0}$:}}
\begin{equation}
\begin{array}{ll}
\nabla_{\bf x} \times {\bf E}_0({\bf x},{\bf y})
+ \nabla_{\bf y} \times {\bf E}_1({\bf x},{\bf y})=i\omega\mu_0 {\bf H}_0({\bf x},{\bf y}) \, , \\[2mm]
\nabla_{\bf x} \times {\bf H}_0({\bf x},{\bf y})
+ \nabla_{\bf y} \times {\bf H}_1({\bf x},{\bf y})=-i\omega\varepsilon({\bf y}) {\bf E}_0({\bf x},{\bf y}) \, , \\[2mm]
\nabla_{\bf x} \cdot \big[ \varepsilon({\bf y}) {\bf E}_0({\bf x},{\bf y}) \big] 
+ \nabla_{\bf y} \cdot \big[ \varepsilon({\bf y}) {\bf E}_1({\bf x},{\bf y}) \big] =0 \, , \\[2mm]
\nabla_{\bf x} \cdot  {\bf H}_0({\bf x},{\bf y})
+ \nabla_{\bf y} \cdot  {\bf H}_1({\bf x},{\bf y})=0 \, . \\
\end{array}
\label{eq-order1}
\end{equation}
The first two lines of the equation above imply:
\begin{equation}
\begin{array}{ll}
\nabla_{\bf y} \times \nabla_{\bf y} \times 
{\bf E}_1({\bf x},{\bf y})=
- \nabla_{\bf y} \times \nabla_{\bf x} \times 
{\bf E}_0({\bf x},{\bf y}) \, , \\[2mm]
\nabla_{\bf y} \times \nabla_{\bf y} \times 
{\bf H}_1({\bf x},{\bf y}) = 
-i\omega \nabla_{\bf y} \times \varepsilon({\bf y}) {\bf E}_0({\bf x},{\bf y}) \, .
\end{array}
\label{eq-order1a}
\end{equation}
and the last two lines imply:
\begin{equation}
\begin{array}{ll}
\nabla_{\bf y} \cdot \big[ \varepsilon({\bf y}) {\bf E}_1({\bf x},{\bf y}) \big] = 
- \nabla_{\bf x} \cdot \big[ \varepsilon({\bf y}) {\bf E}_0({\bf x},{\bf y}) \big] \, , \\[2mm]
\nabla_{\bf y} \cdot  {\bf H}_1({\bf x},{\bf y})=
- \nabla_{\bf x} \cdot  {\bf H}_0({\bf x},{\bf y})\, .
\end{array}
\label{eq-order1b}
\end{equation}


\textcolor{black}{
Applying the mean operator $\bla \cdot \bra$ over the periodic cell $Y={[0,1]}^3$, and invoking the generalized Stokes' theorem applied over $Y$ (a torus, so a manifold without boundary), the first two lines in the set of equations (\ref{eq-order1}) of order $\eta^0$ above imply
\begin{equation}
\begin{array}{ll}
\nabla_{\bf x} \times \bla {\bf E}_0({\bf x},{\bf y}) \bra
=i\omega\mu_0 \bla {\bf H}_0({\bf x},{\bf y}) \bra \, , \\[2mm]
\nabla_{\bf x} \times \bla{\bf H}_0({\bf x},{\bf y}) \bra
=-i\omega \bla\varepsilon({\bf y}) {\bf E}_0({\bf x},{\bf y}) \bra \, . \\[2mm]
\end{array}
\label{maxwellhom0}
\end{equation}
}

{\bg{The third line in equation (\ref{eq-order0})}} provides us with the homogenized permittivity simply by assuming the leading order term of the asymptotic expansion (\ref{ansatzE}) has the following form
{\color{black}{\cite{guenneau2000}}}
\begin{equation}
\begin{array}{ll}
{\bf E}_0({\bf x},{\bf y}) & = \bla {\bf E}_0 \bra ({\bf x}) + \nabla_{\bf y} 
{\color{black}{ {\bf V^{(0)}} }} ({\bf y})\cdot \bla {\bf E}_0 \bra ({\bf x}) 
\label{E-order0} 
\end{array} 
\end{equation}
This form obviously fulfills the partial differential equation {\bg{given by the first line in (\ref{eq-order0}), since the composition of the curl operator $\nabla_{\bf y}\times$ with $\nabla_{\bf y}$}} vanishes and plugging it in the third eqaution of (\ref{eq-order0}) leads to the annex problem {\color{black}{\cite{guenneau2000}}}
\begin{equation}
\nabla_{\bf y} \cdot \left( \varepsilon({\bf y}) \big[ \bla {\bf E}_0 \bra ({\bf x}) + \nabla_{\bf y} {\color{black}{\bf V^{(0)}}}({\bf y})\cdot \bla {\bf E}_0 \bra({\bf x}) \big] \right) =0 
\end{equation}
or equivalently
\begin{equation}
\nabla_{\bf y} \cdot \left(\varepsilon({\bf y}) \big[ {\bf e}_i+ \nabla_{\bf y} {\color{black}{V^{(0)}_i}}({\bf y}) \big] \right) = 0 \; , \; i=1,2,3 \; ,
\label{leadingannex}
\end{equation}
where $\bla \cdot \bra$ denotes the mean operator over the periodic cell $Y={[0,1]}^3$ and {\bg{${\bf V^{(0)}}$}} is a rank-1 tensor such that {\bg{$\bla {\bf V^{(0)}} \bra ={\bf 0}$}} and {\color{black}{and $\{{\bf e}_1,{\bf e}_2,{\bf e}_3\}$ is the canonical basis in Euclidean space $\R^3$.}} 

We now apply the mean operator to {\bg{the third line in the equation (\ref{eq-order1}) of order $\eta^0$:}}
\begin{equation}
\nabla_{\bf x} \cdot \bla \varepsilon({\bf y}) {\bf E}_0({\bf x},{\bf y}) \bra
+ \bla \nabla_{\bf y} \cdot \big[ \varepsilon({\bf y}) {\bf E}_1({\bf x},{\bf y})\big] \bra =0 \, ,
\end{equation}
{\bg{where, thanks to the divergence theorem, the second term reduces to the integral over the boundary $\partial Y$ of the unit cell $Y$, }}
$ \int_{\partial Y} \varepsilon {\bf E}_1 \cdot{\bf n} ds $. Furthermore, noting that ${\bf E}_1({\bf x},{\bf y})$ is periodic over $Y$ and the normal ${\bf n}$ is anti-periodic, this term vanishes. Using the ansatz (\ref{E-order0}), we get
\begin{equation}
\nabla_{\bf x} \cdot \left( \bla \varepsilon({\bf y}) \big[ {\color{black}{I}}+\nabla_{\bf y} {\color{black}{{\bf V}^{(0)}}} \big] \bra \cdot \bla {\bf E}_0 \bra ({\bf x})\right)=0
\label{hom0}
\end{equation}
which {\bg{provides the homogenized equation for the electric field at order $0$, with effective rank-2 anisotropic permittivity tensor with coefficients}}
\begin{equation}
{\color{black}{
\varepsilon_{hom,ij}^{(0)}}}= \bla \varepsilon({\bf y}) \big[ {\delta}_{ij}+{\color{black}{\frac{\partial{V^{(0)}_i}}{\partial y_j}}} \big] \bra  \; , \; i=1,2,3 \; ,
\label{leadingpermittivity}
\end{equation}
where $\delta_{ij}$ is the Kronecker symbol.

{\bg{
As for the leading order term of the asymptotic expansion (\ref{ansatzH}), it is given by 
\begin{equation}
\begin{array}{ll}
{\bf H}_0({\bf x},{\bf y}) & = \bla {\bf H}_0 \bra ({\bf x}) \, ,
\label{H-order0} 
\end{array} 
\end{equation}
}}
since, from (\ref{eq-order0}), $\nabla_{\bf y} \times  {\bf H}_0({\bf x},{\bf y})=0$ and 
$\nabla_{\bf y} \cdot  {\bf H}_0({\bf x},{\bf y})=0$.

\textcolor{black}{
We note that from (\ref{maxwellhom0}) and (\ref{hom0}) we get the homogenized Maxwell's equations at the leading order:
\begin{equation}
\begin{array}{ll}
\nabla_{\bf x} \times {\bf E}_{hom}^{(0)}
=i\omega\mu_0 {\bf H}_{hom}^{(0)} ({\bf x}) \, , \\[2mm]
\nabla_{\bf x} \times {\bf H}_{hom}^{(0)} ({\bf x})
=-i\omega \varepsilon_{hom}^{(0)} {\bf E}_{hom}^{(0)} \, , \\[2mm]
\end{array}
\label{maxwellhom0bis}
\end{equation}
with
\begin{equation}
\begin{array}{l}
    {\bf E}_{hom}^{(0)} ({\bf x}) = \bla {\bf E}_0 \bra ({\bf x})  \, , \\[2mm]
    {\bf H}_{hom}^{(0)} ({\bf x}) = \bla {\bf H}_0 \bra ({\bf x}) \, .  \\[2mm]
\end{array}
\end{equation}
}

\subsection{Effective parameters at higher orders}

{\bg{Let us now proceed with the next terms ${\bf E}_1({\bf x},{\bf y})$ and ${\bf H}_1({\bf x},{\bf y})$ in 
the expansion of the fields (\ref{ansatzE}) and (\ref{ansatzH}). The equation (\ref{eq-order1}) shows that 
each of these first-order term is the solution to a linear equation with two forcing terms generated by 
the zero-order components ${\bf E}_0({\bf x},{\bf y})$ and ${\bf H}_0({\bf x},{\bf y})$. 
Hence the first order introduces a coupling between the electric and magnetic fields. 

Let us now proceed with the terms 
of order $\eta^{1}$ following the equation (\ref{eq-order1}) in a hierarchy of equations {\bg{in powers of $\eta$.}}
\textcolor{black}{We can see that all terms ${\bf H}_i$ are divergence free, so all the effective properties will come from $\nabla\cdot(\varepsilon^{-1}{\bf E})=0$. More precisely,}
\begin{equation}
\begin{array}{ll}
\nabla_{\bf x} \times {\bf E}_1({\bf x},{\bf y})
+ \nabla_{\bf y} \times {\bf E}_2({\bf x},{\bf y})=i\omega\mu_0 {\bf H}_1({\bf x},{\bf y}) \, , \\[2mm]
\nabla_{\bf x} \times {\bf H}_1({\bf x},{\bf y})
+ \nabla_{\bf y} \times {\bf H}_2({\bf x},{\bf y})=-i\omega\varepsilon({\bf y}) {\bf E}_1({\bf x},{\bf y}) \, , \\[2mm]
\nabla_{\bf x} \cdot \big[ \varepsilon({\bf y}) {\bf E}_1({\bf x},{\bf y})\big]
+ \nabla_{\bf y} \cdot \big[\varepsilon({\bf y}) {\bf E}_2({\bf x},{\bf y})\big]=0 \, , \\[2mm]
\nabla_{\bf x} \cdot {\bf H}_1({\bf x},{\bf y})
+ \nabla_{\bf y} \cdot  {\bf H}_2({\bf x},{\bf y}) =0 \, .\\
\end{array}
\label{eq-order2}
\end{equation}
\textcolor{black}{
Applying the mean operator $\bla \cdot \bra$ over the periodic cell $Y={[0,1]}^3$, and invoking the generalized Stokes' theorem applied over $Y$ (a torus, so a manifold without boundary), the first two lines in the set of equations (\ref{eq-order2}) of order $\eta^1$ above imply
\begin{equation}
\begin{array}{ll}
\nabla_{\bf x} \times \bla {\bf E}_1({\bf x},{\bf y}) \bra
=i\omega\mu_0 \bla {\bf H}_1({\bf x},{\bf y}) \bra \, , \\[2mm]
\nabla_{\bf x} \times \bla{\bf H}_1({\bf x},{\bf y}) \bra
=-i\omega \bla\varepsilon({\bf y}) {\bf E}_1({\bf x},{\bf y}) \bra \, . \\[2mm]
\end{array}
\end{equation}
}

Applying the mean operator $\bla \cdot \bra$ over the periodic cell $Y={[0,1]}^3$, and making use of the divergence theorem applied over $Y$, the last two lines in the set of equations (\ref{eq-order2}) of order $\eta^1$ imply
\begin{equation}
\begin{array}{ll}
\nabla_{\bf x} \cdot \bla \varepsilon({\bf y}) {\bf E}_1({\bf x},{\bf y}) \bra =0 \, , \\[2mm] 
\nabla_{\bf x} \cdot \bla {\bf H}_1({\bf x},{\bf y}) \bra =0 \, ,
\label{divfree-order2}
\end{array}
\end{equation}
and, in addition with the leading order, lead to
\begin{equation}
\begin{array}{ll}
\nabla_{\bf x} \cdot \bla \varepsilon({\bf y}) \big[ {\bf E}_0({\bf x},{\bf y}) + 
\eta {\bf E}_1({\bf x},{\bf y}) \big] \bra =0 \, , \\[2mm] 
\nabla_{\bf x} \cdot \bla \mu_0 \big[ {\bf H}_0({\bf x},{\bf y}) + \eta 
{\bf H}_1({\bf x},{\bf y}) \big] \bra =0 \, .
\label{const-order2}
\end{array}
\end{equation}
These two equations are the starting point to derive the effective constitutive equations at the order $\eta^{1}$. 
Indeed, using that 
${\bf H}_0({\bf x},{\bf y}) = \bla {\bf H}_0 \bra ({\bf x})$ is independent of the variable 
${\bf y}$, we have $\nabla_{\bf x} \cdot {\bf H}_0({\bf x},{\bf y}) = 
\nabla_{\bf x} \cdot \bla {\bf H}_0 \bra ({\bf x}) = 0$ and $\nabla_{\bf y} \times 
\nabla_{\bf x} \times {\bf H}_0({\bf x},{\bf y}) = \nabla_{\bf y} \times 
\nabla_{\bf x} \times \bla {\bf H}_0 \bra ({\bf x}) = 0$. Therefore, the second and fourth 
lines of equation (\ref{eq-order1}) imply
\begin{equation}
\begin{array}{ll}
\nabla_{\bf y} \times \nabla_{\bf y} \times {\bf H}_1({\bf x},{\bf y})=-i\omega
\nabla_{\bf y} \times [\varepsilon({\bf y}) {\bf E}_0({\bf x},{\bf y})] \, , \\[2mm]
\nabla_{\bf y} \cdot  {\bf H}_1({\bf x},{\bf y})=0 \, . \\
\end{array}
\label{eq-H1}
\end{equation}
From the first line in (\ref{eq-H1}) and (\ref{E-order0})
\begin{equation}
\begin{array}{ll}
\nabla_{\bf y} \times \nabla_{\bf y} \times {\bf H}_1({\bf x},{\bf y})
&=-i\omega
\nabla_{\bf y} \times [\varepsilon({\bf y}) \left(\bla {\bf E}_0 \bra ({\bf x}) + \nabla_{\bf y} 
{\color{black}{ {\bf V^{(0)}} }} ({\bf y})\cdot \bla {\bf E}_0 \bra ({\bf x})\right) ] \\[2mm]
&=-i\omega
\nabla_{\bf y} \varepsilon({\bf y}) \times \bla {\bf E}_0 \bra ({\bf x})
\\
\end{array}
\label{eq-H1}
\end{equation}

We can write
\begin{equation}
\begin{array}{ll}
{\bf H}_1({\bf x},{\bf y}) & = \bla {\bf H}_1 \bra ({\bf x})
+
{\bf W}_1({\bf y}) \times \bla {\bf E}_0 \bra ({\bf x})
\label{H-order1new} 
\end{array} 
\end{equation}
where ${\bf W}_1({\bf y})$ is a rank-1 tensor solution of
\begin{equation}
\begin{array}{ll}
\nabla_{\bf y} \times \nabla_{\bf y} \times {\bf W}_1({\bf y})
=-i\omega
\nabla_{\bf y} \varepsilon({\bf y})
\end{array}
\label{eq-H1anew}
\end{equation}
We note that since $\nabla_{\bf y} \cdot  {\bf H}_1({\bf x},{\bf y})=0$, one has that
$\nabla_{\bf y} \cdot ({\bf W}_1({\bf y}) \times \bla {\bf E}_0 \bra ({\bf x}))
=0$.

\color{black}{}Similarly, the first and third 
lines of equation (\ref{eq-order1}) imply
\begin{equation}
\begin{array}{ll}
\nabla_{\bf y} \times \nabla_{\bf y} \times {\bf E}_1({\bf x},{\bf y})=-
\nabla_{\bf y}\times\nabla_{\bf x} \times {\bf E}_0({\bf x},{\bf y}) \, , \\[2mm]
\nabla_{\bf y} \cdot \big[ \varepsilon({\bf y}) {\bf E}_1({\bf x},{\bf y}) \big] =
-\varepsilon({\bf y}) \nabla_{\bf x} \cdot \big[ {\bf E}_0({\bf x},{\bf y}) \big] \, , \\
\end{array}
\label{eq-H1toE1}
\end{equation}
since $\nabla_{\bf y} \times {\bf H}_0({\bf x},{\bf y})=\nabla_{\bf y} \times {\bf H}_0({\bf x})={\bf 0}$ and the expression for 
${\bf E}_1({\bf x},{\bf y})$ ensues.

\end{document}